\documentclass[twocolumn,showpacs,aps,floatfix]{revtex4}

\usepackage{psfig}
\usepackage{graphicx}
\usepackage{dcolumn}
\usepackage{bm}
\usepackage{amsmath}

\newcommand{\ud}{\mathrm{d}}

\newcommand{\etal}{{\em et al.}}

\newcommand{\tu}[1]{\mathrm{#1}}

\begin{document}
\title{Cosmogenic Neutrinos from Cosmic Ray Interactions with 
Extragalactic Infrared Photons}

\author{Todor Stanev}
\email{stanev@muon.bartol.udel.edu}
\affiliation{Bartol Research Institute, University of Delaware, Newark, 
DE 19716, USA}
\author{Daniel De Marco}
\email{ddm@bartol.udel.edu}
\affiliation{Bartol Research Institute, University of Delaware, Newark, 
DE 19716, USA}
\author{F.W.~Stecker}
\email{Floyd.W.Stecker@nasa.gov}
\affiliation{NASA Goddard Space Flight Center}

\date{\today}

\begin{abstract}
 We discuss the production of cosmogenic neutrinos on extragalactic 
 infrared photons in a model of its cosmological evolution. The relative
 importance
 of these infrared photons as a target for proton interactions
 is significant, especially in the case of steep injection spectra of
 the ultrahigh energy cosmic rays. For an E$^{-2.5}$ cosmic ray
 injection spectrum, for example, the event rate of neutrinos of
 energy above 1 PeV is more than doubled.  
\end{abstract}

\maketitle

\section{Introduction}
 The assumption that the ultra high energy cosmic rays (UHECR) are
 nuclei (presumed here to be protons) accelerated in powerful extragalactic
 sources provides a natural connection between these particles and
 ultra high energy neutrinos. This was first realized by
 Berezinsky and Zatsepin~\cite{BerZat69} soon after the introduction of
 the GZK effect~\cite{GZK}. The GZK effect is the modification of the
 UHE proton spectrum from energy losses by photoproduction
 interactions with the 2.7K microwave background radiation (MBR).
 In the case of isotropic and homogeneous distribution of
 UHE cosmic ray sources, the GZK effect leads to a cut-off 
 of the cosmic ray spectrum below 10$^{20}$ eV.  
 The charged mesons generated in these interactions initiate
 a decay chain that results in neutrinos. Since the mesons and 
 muons do not lose energy before decay, the high energy end of 
 the spectrum of these neutrinos follows the injection spectrum
 of UHECR, while below the interaction threshold it is flat
 \cite{Stecker73}, \cite{Stecker79}.
 The neutrinos which are produced by photomeson producing
interactions of UHECR nuclei are sometimes referred to as
{\em cosmogenic neutrinos}. 
 
 Several calculations of of the fluxes of UHE photomeson neutrinos
were published in the 1970s \cite{WTW,Stecker73,BerSmi75,BerZat77,Stecker79},
Hill and Schramm~\cite{HS85,HSW86} used the
 non-detection of such neutrinos to place an upper limit on the cosmological 
evolution of the sources of UHECR.  The problem has been revisited several
 more times~\cite{YT93,PJ96,ESS01}.

 In 2004 Stanev~\cite{TS04} considered interactions of UHECR
with photons of the extragalactic infrared and optical background 
(IRB), pointing out that this process 
 generates non-negligible cosmogenic neutrino fluxes. This suggestion
 was quickly followed by a confirmation in Ref.~\cite{BMM04} 
 which emphasized the importance of the IRB as
 interaction target. This idea was further developed in
 Ref.~\cite{BK05}.

 Ref.~\cite{TS04} gave an estimate of the
 cosmogenic neutrino flux generated in interactions on the IRB, but
 did not account correctly for the cosmological evolution of the
 infrared background. In this paper we perform a calculation using a realistic
empirically based  model of the
 cosmological evolution of the spectral energy distribution of the
extragalactic IR-UV  background given in Ref. \cite{SMS05} which
will be referred to as SMS05. The aim is to estimate correctly the 
role of these extragalactic
photons, particularly the infrared photons which are by far the most
numerous, as targets for UHE proton interactions. 

 The paper is organized as follows: in Section II we discuss the
 model of the infrared background and its cosmological evolution.
 In Section III we describe
 the calculation. Section IV gives the results of the calculation
 and Section V contains the discussion of the results and the
 conclusions from this research.

\section{Cosmological Evolution of the IR-UV  background}

It is  now well known that galaxies  had a brighter past  owing to the
higher rate  of star formation  which took place. Strong  evolution is
supported by many observations  relating IR luminosity to the much
higher star formation rate at $z \sim 1$ and to the recent determination
that  most  Lyman break  galaxies  at $z  \sim  1$  are also  luminous
infrared galaxies. In addition to the evolution
of galaxy luminosity, some increase in galaxy number density is expected
owing to the hierarchical clustering predicted by cold dark matter models.
However, luminosity evolution is the dominant effect and it is difficult
to separate out a component of density evolution.

In order to calculate intergalactic IR photon fluxes and densities
and their evolution over time (or redshift), 
SMS05 performed an empirically based calculation
the SED of the  IBR (infrared background radiation) by using  (1) the
luminosity dependent galaxy spectral energy distributions (SEDs) based on
galaxy observations,
(2) observationally derived  galaxy luminosity distribution functions  (LFs)
and  (3)   the  latest  redshift  dependent  luminosity
evolution  functions, sometimes  referred  to  as  Lilly-Madau plots. 
The SMS05 calculation was an improved version of the work presented
in Refs. \cite{MS98}, \cite{SS98} and \cite{MS01}. 

 The calculation considers two different cosmological evolutions, ${\cal E}(z)$
 {\em baseline} and {\em fast}, of the
 infrared emission of the type
\begin{equation}
  {\cal E}(z) =  \left\{
\begin{array}{lcl}
 (1 + z )^{m}  &  {\rm :}   &  z < z_\tu{flat}\\
 (1 + z_\tu{flat})^{m} &  {\rm :}   & z_\tu{flat} < z < 6\\
  0 & {\rm :} & z >  6
\end{array}\right. 
\label{evo} 
\end {equation}
 The {\em baseline} evolution model is described by
 $m$=3.1 and $z_\tu{flat}$ = 1.3, while the {\em fast} evolution model
 uses $m$=4 and $z_\tu{flat}$ = 1. The infrared emission at $z > z_\tu{flat}$ 
 is constant in both models.
\begin{figure}[htb]
\centerline{\includegraphics[width=83truemm]{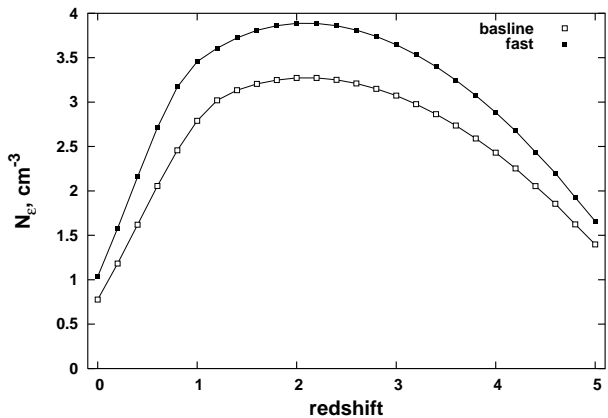}}
  \caption{ Number density of the IRB  at
  different redshifts as calculated by SMS05 \protect\cite{SMS05}.}
\label{fig:irbg_den}
\end{figure}
 Figure~\ref{fig:irbg_den} shows the number density between 
 photon energy of 3.16$\times$10$^{-3}$ and 1 eV in both models.
 The {\em fast} evolution model has higher density in the current
 cosmological epoch as well at the IRB maximum epoch, which is 
 around $z$ = 2. The increase of the total IRB number density
 increases by a factor of about 4 from $z$ = 0 to $z$ = 2 and
 decreases at higher redshifts. One should note, however, that
 the cosmological evolution of the infrared background density
 is much slower than that of MBR since the current IRB density
 is accumulated from the infrared emission of different sources
 since $z$ = 6.   
\begin{figure}[htb]
\centerline{\includegraphics[width=83truemm]{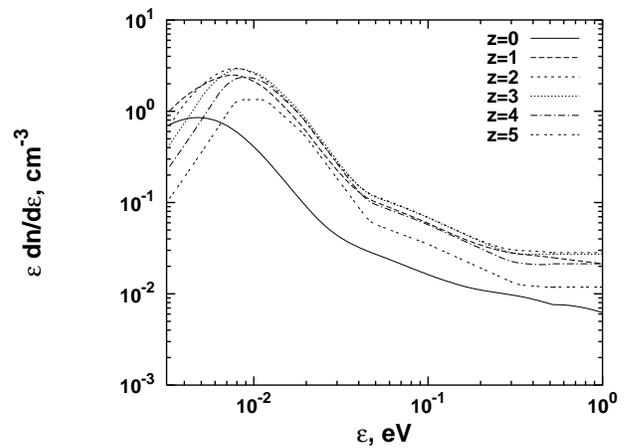}}
\caption{ Number density of the infrared background
 at different redshifts calculated by SMS05 \protect\cite{SMS05}
 in the {\em fast} evolution model.}
\label{fig:irbg_z}
\end{figure}
 Figure~\ref{fig:irbg_z} shows the energy spectrum of the {\em fast} 
 infrared background at redshifts from 0 to 5. One can see both
 the increase of the total photon density as well as the shift
 of the maximum of the emission to higher energy at higher redshifts.
 In terms of photoproduction interactions on IRB this means that
 lower energy cosmic rays will be above the photoproduction 
 threshold at higher redshifts. 

 Both figures above show the number density
 of IRB rather than the usual presentation of the energy density.
 Since we are using the infrared background as a target for cosmic
 ray interactions this is the relevant quantity.

\section{The calculation}

 The calculation was performed in two stages: (1)
 calculation of the neutrino yields from interactions with
 extragalactic infrared photons and (2) a subsequent integration of the
 yields to obtain the cosmogenic neutrino flux from such
 interactions. This approach gives us the flexibility to
 easily obtain the neutrino flux using different parametrizations 
 of the cosmic ray emissivity, injection spectra and 
 cosmological evolution of the cosmic ray sources.
This approach, however, suffers from the fact that since 
 the yields are calculated only on the IRB, they do not 
 account for the fact that high energy protons interact 
 mainly with the much more numerous MBR photons.

\subsection{Calculation of the Neutrino Yields}

 The neutrino yields as a function of the proton energy E$_p$, 
 neutrino energy E$_\nu$ and the redshift, $z$, were calculated
 using the IRB spectra at different
 cosmological epochs, {\it i.e.}, as a function of redshift, 
that were provided by the authors of 
Ref.~\cite{SMS05}. Each of the yield calculations was performed
for proper distances corresponding to $\Delta z$=0.2 
using an  $\Omega_M$=0.3, $\Omega_\Lambda$=0.7 cosmology as
\begin{equation}
  D(z) = \frac{c}{H_0}\int_{z_\tu{min}}^{z_\tu{max}}
 \frac{1}{1 + z} \left[\Omega_M (1+z)^3 + \Omega_\Lambda \right]^{-1/2}
\label{dddz}
\end{equation} 
 The IRB is considered to be constant during each cosmological epoch
of duration $\Delta z$=0.2.

 In this way the matrix element corresponding to $\ud t/\ud z$
 dependence was accounted for in the yields. The yields
 were calculated with the code used in Ref.~\cite{ESS01}
 and the photoproduction interaction code SOPHIA~\cite{SOPHIA}.
 All generated neutrinos are redshifted by the code to the
 end of the $\Delta z$ epoch. The yields were 
 calculated for redshifts $ 0 < z < 5$ and for cosmic ray
 energies above 10$^{18}$ eV in ten logarithmic bins per
 decade of energy.

 \begin{figure}[htb]
\centerline{\includegraphics[width=83truemm]{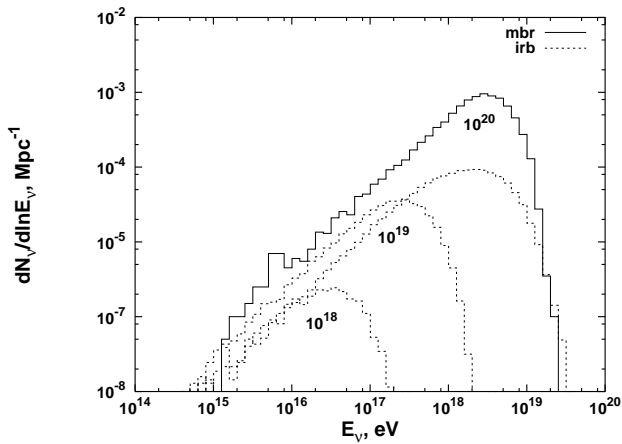}}
  \caption{ Neutrino yields for  10$^{20}$ eV protons 
interacting with both MBR photons and for 10$^{18}$ eV, 10$^{19}$ eV 
and 10$^{20}$ eV protons interacting with IRB photons at 
$z = 0$, given for protons traveling a distance of 1 Mpc.}
\label{fig:yield}
\end{figure}

 Fig.~\ref{fig:yield} compares the $\nu_\mu$ yields for UHE protons
 traveling a distance of 1 Mpc and interacting with MBR and IRB photons.
 The yield for 10$^{20}$ eV
 protons interacting with IRB photons is about a factor of 10 lower than 
that for MBR interactions. This difference is much smaller than the 
 ratio of the MBR and IRB total densities, and demonstrates that 
 10$^{20}$ eV protons interact mainly with photons in the higher frequency
Wien tail of the 2.7K MBR spectrum.

 Protons of energy 10$^{19}$ eV do not interact at $z$=0 with MBR photons,
 but they 
 readily interact and produce neutrinos by interactions with IRB photons, 
 as do protons
 of energy 10$^{18}$ eV. Even protons of energy 10$^{17}$ eV
 occasionally interact with IRB photons, but their contribution is very
 small and is neglected in this calculation. Even for $E^{-2}$
 cosmic ray spectra, the smaller 10$^{19}$ and 10$^{18}$
 eV yields are multiplied by the much higher flux of cosmic rays with 
 such energies. This is the basis of the significant neutrino
 (and $\gamma$-ray) production from UHECR-IRB interactions.

\subsection{Integration of the Yields}

 The second phase of the calculation requires the parametrization of
 the redshift evolution of the emissivity of cosmic ray sources, 
 and the form of the cosmic ray injection spectrum. We assume a cosmic ray 
 injection spectrum of the power-law form 
 $\ud N/\ud E_p = A E_p^{-(\gamma + 1)} \exp(-E_p/E_\tu{max})$
 with $E_\tu{max}$ = 10$^{21.5}$ eV. 

 We consider here two empirically based models for the evolution of
 UHECR power with redshift, {\it viz.}, (1) one based on the redshift
 evolution of the star formation rate that was used in the calculation
 of the infrared background in SMS05, and (2) the other based on the
 redshift evolution of flat spectrum radio sources, given 
 as an analytic approximation in Ref. \cite{DP90}.  
 We use the {\em fast} evolution model from SMS05 since it is more
 consistent with the new observations of the Spitzer
 telescope~\cite{Spitzer2,Spitzer1}.

 We normalize the cosmic ray energy flux at $E_p$ = 10$^{19}$ eV to
 $E_p dN_p/E_p$ = 2.5$\times$10$^{-18}$ cm$^{-2}$s$^{-1}$sr$^{-1}$
 \cite{W95,stsc05}.  
 Since the calculation is extended to energies 
 below 10$^{19}$ eV, the code uses the cosmic ray 
 flux at 10$^{19}$ eV to calculate the injection spectra at lower 
 and higher energy.  
 Therefore, the cosmic ray emissivity above 10$^{18}$ eV
 depends on the injection spectrum. 
 The injection spectrum itself is used as a free parameter in order to
 study its influence on the cosmogenic neutrino spectrum.

\begin{figure}[htb]
 \centerline{\includegraphics[width=83truemm]{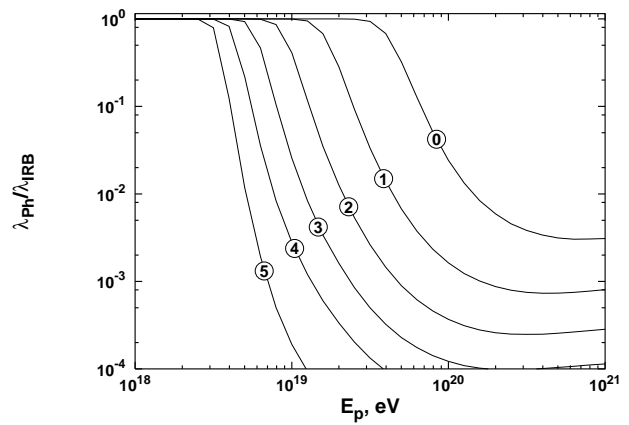}}
 \caption{Fraction of the total cosmic ray flux used in the integration
 of the neutrino yields from interactions in the IRB. The different
 lines correspond to fractions at different redshifts as indicated by
 the numbers in the plot.}\label{fig:frac}
\end{figure}

 The integration procedure also has to account for the modification 
 of the cosmic ray spectrum owing to interactions with MBR photons. 
 This was done in two crude, but reasonable, ways. The first one is 
 the introduction of a high energy cutoff of the spectrum as a 
 function of the redshift. The second one, which is used in 
 the results presented below, is to weight the cosmic ray injection
 spectrum with the interaction length $\lambda_\tu{IRB}$ on the  
 infrared background radiation. The cosmic rays interacting in 
 the IBR used in the integration of the yields are $F_\tu{CR} \;
 \lambda_\tu{IRB}/\lambda_\tu{tot}$, where $\lambda_\tu{tot}$ is the
 interaction length in the total IRB and MBR fields. The fraction of the
 cosmic ray flux used in the integration procedure is shown in
 Fig.~\ref{fig:frac}.
 If one arbitrarily determines the high energy cutoff of the 
 cosmic ray energy spectrum as the energy at which only 10 per cent of
 the cosmic rays interact in the IRB, it would be 7$\times$10$^{19}$ eV
 at $z$ = 0 compared to 2.5$\times$10$^{19}$ and 4$\times$10$^{18}$ eV
 at $z$ = 1 and 5.
 
 Since the yields include the
 $\ud t/\ud z$ factor the integration becomes very simple, {\it viz.}:
\begin{align}
 \nonumber & \ud N_\nu^i/\ud E_\nu = 
 \int_0^5 \ud z \; \times \\
 & \times \int^{E_c} \frac{\ud N_p}{\ud E_p}\, {\cal E}(z) \,
 Y^i \left[(1+z)E_\nu; E_p ,z \right] \ud E_p \; ,
\end{align}
 where the index $i$ indicates the neutrino flavor.

\section{Results}

\begin{figure}[thb]
 \centerline{\includegraphics[width=82truemm]{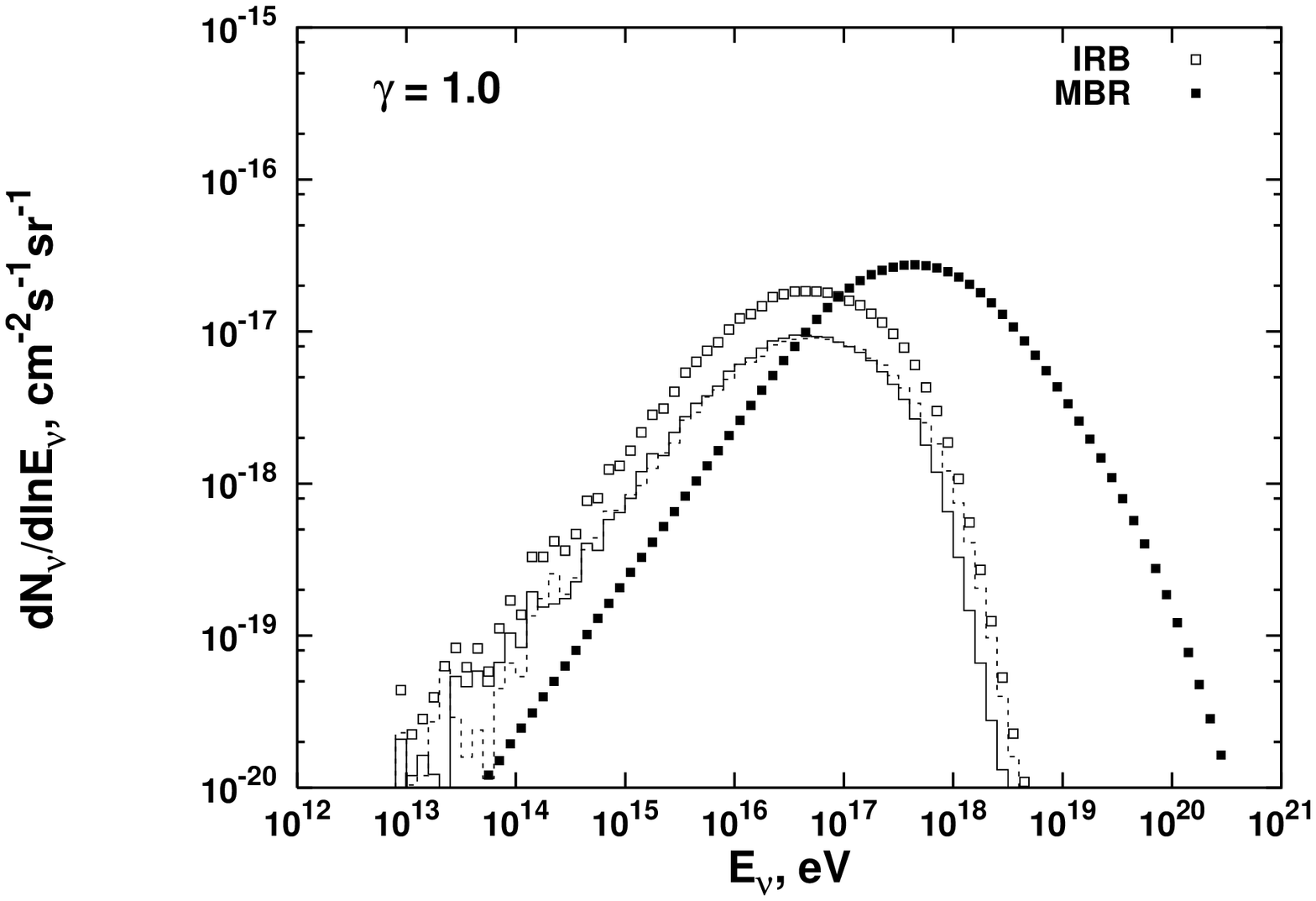}}
 \centerline{\includegraphics[width=82truemm]{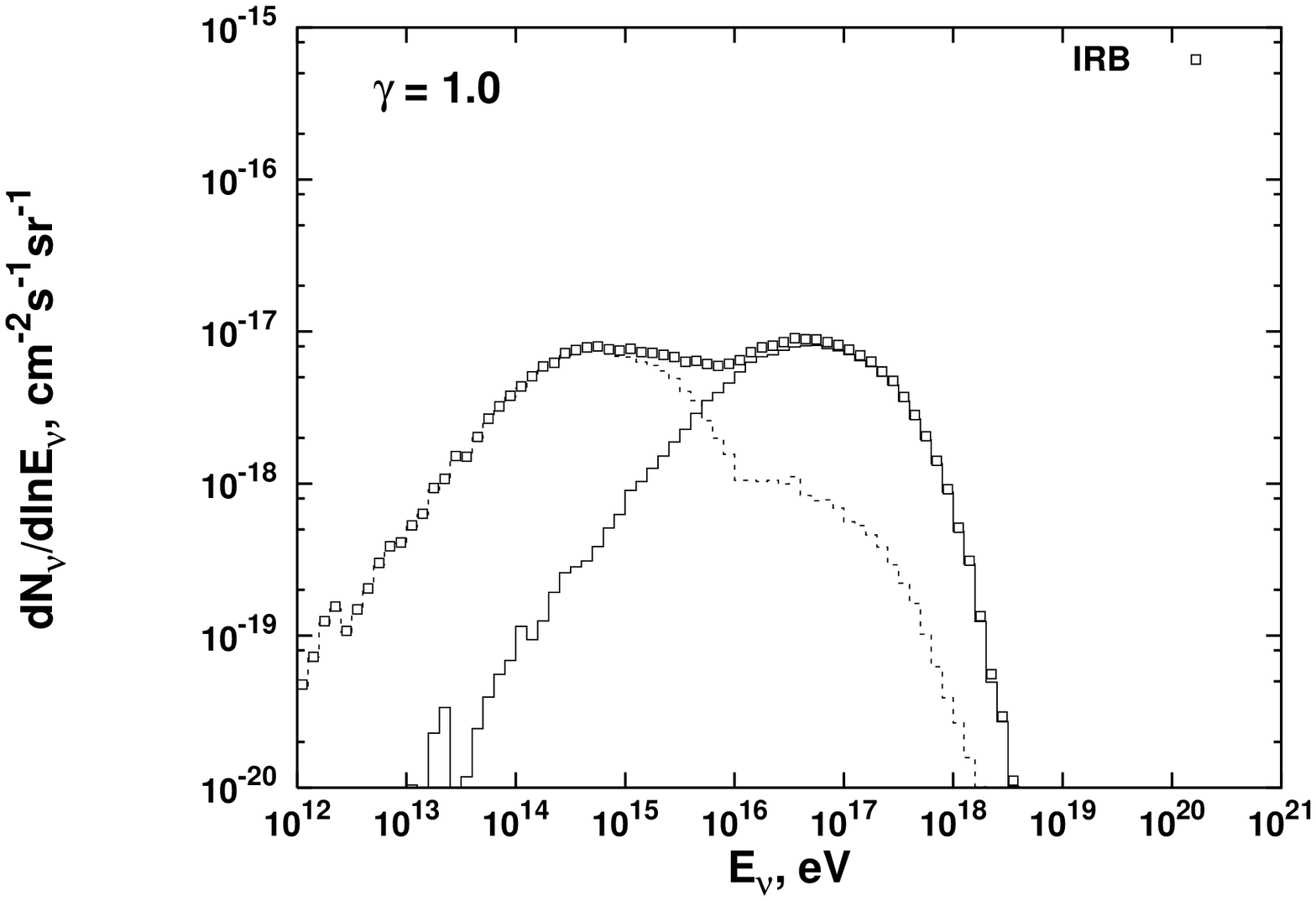}}
 \centerline{\includegraphics[width=82truemm]{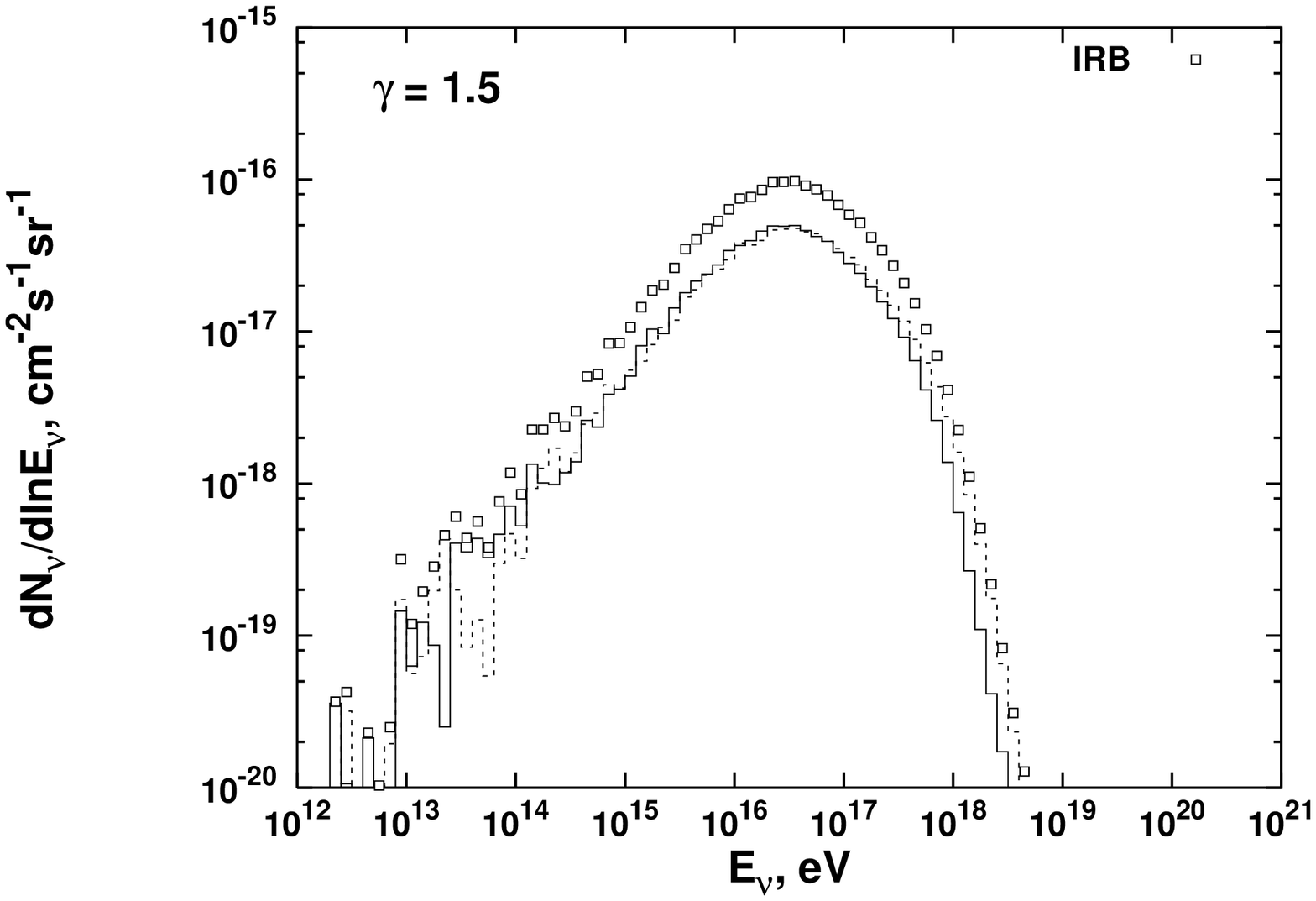}}
 \caption{ Top panel: $\nu_\mu$ (solid) and $\bar{\nu}_\mu$ (dashed)
  spectra generated by interactions with  IRB photons. Their sum 
  (open squares) 
  is compared to those generated by interactions with MBR (MBR) 
  photons (full squares) for \protect$\gamma$=1 
  assuming {\em fast} evolution of the cosmic ray source 
  emissivity.
  Middle panel: $\nu_e$ (solid) and $\bar{\nu}_e$ 
  (dash) spectra for injection spectra as in the top panel. 
  Their sum is shown with open squares. Bottom panel: $\nu_\mu +
  \bar{\nu}_\mu$ spectra for injection spectrum with $\gamma$=1.5. }
\label{fig:fluxnu1}
\end{figure}

 The results of the integration are shown in Fig.~\ref{fig:fluxnu1}.
 The top panel of the figures compares the fluxes of cosmogenic
 $\nu_\mu + \bar{\nu}_\mu$ neutrinos generated by interactions with
 MBR photons (histogram) with those generated by interactions with IRB
 photons (squares), assuming a $\gamma$ = 1 UHECR injection spectrum
 with the {\em fast} evolution of the emissivity of the cosmic ray
 sources. All panels of Fig.~\ref{fig:fluxnu1} are calculated with the
 same cosmological evolution. Using the {\it baseline} evolution model
 will give neutrino fluxes which are about 25-30\% lower.
  
 The peak flux of the IRB-generated neutrinos is not much lower than that 
 of the MBR-generated ones, {\it i.e.}, 1.7$\times$10$^{-17}$ compared to
 2.2$\times$10$^{-17}$ cm$^{-2}$s$^{-1}$sr$^{-1}$. The peak is, 
 however, shifted to lower E$_\nu$ by about a factor of 3.
 The main reason for that shift is the contribution of 
 protons of energy below 3$\times$10$^{19}$ eV to the
 neutrino production.
 The IRB-generated neutrino flux is also depleted at energies above 
 10$^{19}$ eV. This is because
 protons of energy above 5$\times$10$^{19}$ eV very rarely interact
 with IRB photons before they lose their energy in MBR interactions.
 At energies below the peak
 the IRB-generated neutrino flux is
 somewhat flatter than the MBR one, although the statistical
 uncertainty of the calculation does not allow us to make a 
 quantitative statement regarding this.

 The middle panel of the figure shows the IRB-generated fluxes of 
 $\nu_{e}$'s and  $\bar{\nu}_e$'s assuming an $E^{-2}$ injection 
 spectrum. 
 The electron neutrino flux peaks at the same energy as 
 the muon neutrino one. 
 The $\bar{\nu}_e$ flux, which is due mostly to neutron decay neutrinos, 
 is  shifted and widened at its lower energy end. The dip between 
 the $\nu_e$ and $\bar{\nu}_e$ 
 peaks is not as deep as it is in the MBR neutrino case. The
 reason for that is that the $\nu_e$ peak is somewhat wider
 at energies below the peak. 
 The bottom panel of Fig.~\ref{fig:fluxnu1} shows the fluxes
 of IRB-generated $\nu_\mu+\bar{\nu}_\mu$ assuming a steeper 
 injection spectrum
 $\gamma$ = 1.5 and $m$ = 3.1. There are two main differences from
 the $\gamma$=1 case. The peak of the IRB-generated neutrino flux 
 is higher by almost a factor of 3 (6.5$\times$10$^{-17}$ in 
 the same units) than for the MBR-generated neutrinos and this peak
 and is shifted down in energy by a factor of $\sim 3$ to
 $\sim 10^{16.5}$ eV. This is caused by the higher flux of
 protons of energy below the high energy cutoff. 
 In the case of the MBR-generated neutrinos,
 the general effect is not as strong but is reversed; the
 steeper proton spectrum results in a lower flux.

\begin{figure}[htb]
\centerline{\includegraphics[width=83truemm]{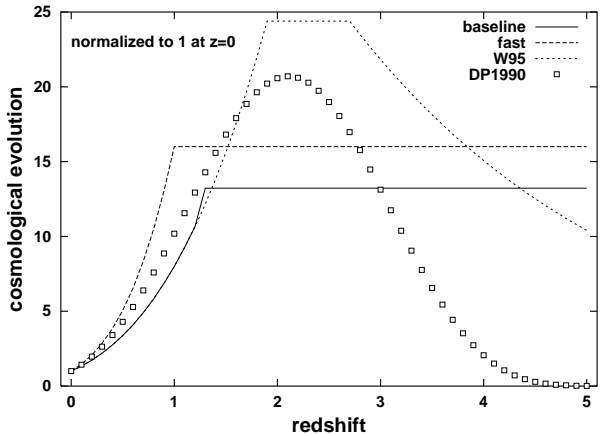}}   
 \caption{ Four different models for the cosmological evolution
 of the cosmic ray sources  (see text for description and 
 references).
 }
\label{fig:cosm}
\end{figure}

 Because of the very strong dependence of the flux of
 cosmogenic neutrinos on the cosmological evolution of 
 the cosmic ray sources~\cite{ss05}, we investigated
 this dependence further. Fig.~\ref{fig:cosm} compares the 
 {\em baseline} and fast cosmological evolutions of 
 Ref.~\cite{SMS05} to these of Refs.~\cite{W95,DP90}.
 All evolution models are normalized to 1 at present, {\it i.e.}
 for $z$ = 0. 

 The evolution taken from Ref.~\cite{W95}, that was used for
 calculations of the cosmogenic neutrino flux from interactions
 in the MBR~\cite{ESS01}, gives a UHECR emissivity which is
 about 60\% higher at redshift of 2  than the average of the
 models of SMS05.   The cosmological evolution of the flat spectrum
 radio galaxies~\cite{DP90} has an intermediate redshift evolution;
 it is faster than $m$ = 3 and slower than $m$ = 4 below $z$ = 1
 and peaks at about $z$ = 2. It is also distinguished by its rapid
 decrease in  emissivity at $z > 3$. 

\begin{figure}[htb]
 \centerline{\includegraphics[width=83truemm]{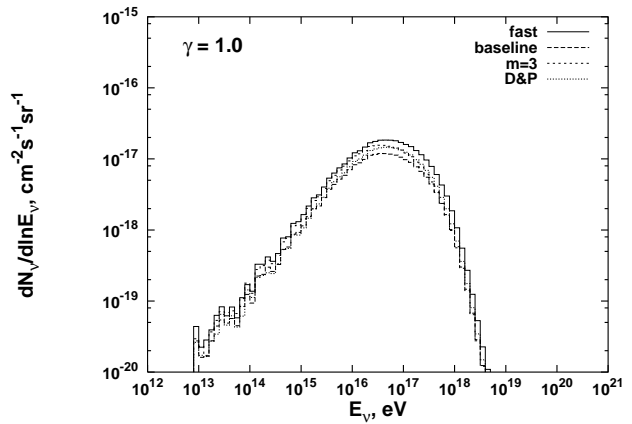}}   
 \caption{Cosmogenic neutrino fluxes for four
 different models of the cosmological evolution 
 of the cosmic ray sources - see text.
 }
\label{fig:cosm_nu}
\end{figure}

 Figure~\ref{fig:cosm_nu} 
 compares the cosmogenic neutrino fluxes of $\nu_\mu + \bar{\nu}_\mu$
 generated by the {\em baseline} and {\em fast} models of SMS05 with the 
 $m$ = 3 model used in Ref.~\cite{ESS01} and that of Ref.~\cite{DP90}.
 The difference in the calculated fluxes is actually quite small,
 compared to all other uncertainties of the calculation. The {\em fast}
 and the {\em baseline} models bracket from above and from below the
 fluxes of cosmogenic neutrinos from interactions in the IRB, while 
 the other two models fall between these two. 
 The main reason for the small differences is the $\ud z/\ud t$ matrix
 element that decreases the contribution of higher redshifts because
 in the cosmological integration the emissivity is multiplied by
 the smaller time intervals involved at higher redshifts. 
 
  In the case where UHECR luminosity evolution is 
 assumed to be proportional to the redshift distribution of flat spectrum 
 radio galaxies as given in Ref.~\cite{DP90}, the neutrino spectra peak
 at a somewhat higher energy because of the relatively small UHECR
 emissivity at the higher redshifts.

\begin{figure}[htb]
\centerline{\includegraphics[width=83truemm]{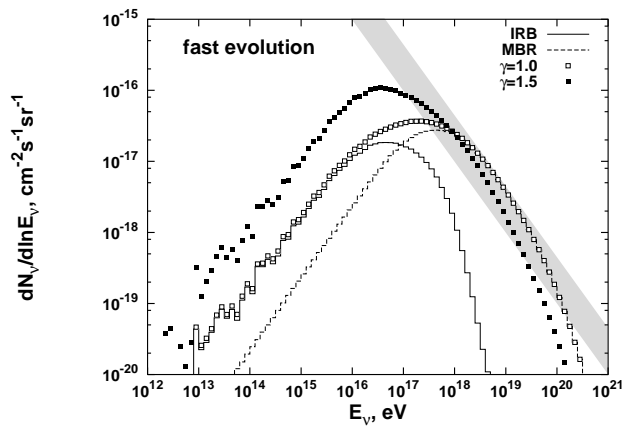}}   
 \caption{Total muon  $\nu_\mu + \bar{\nu}_\mu$ fluxes
 for $\gamma$ = 1.0 (empty squares) and 1.5 (full squares)
 calculated with {\em fast} cosmic ray source evolution.
 The shaded area represents the W\&B~\protect\cite{WB99} upper
 bound on astrophysical source neutrinos.}
\label{fig:fluxnu2}
\end{figure}

 In Fig.~\ref{fig:fluxnu2} we present the total fluxes of cosmogenic
 neutrinos from interactions in the MBR and IRB for injection spectra 
 with $\gamma$ = 1.0 and 1.5 and for {\em fast} cosmological evolution
 of the  cosmic ray sources as in Ref.~\cite{SMS05}.
 
 For a relatively flat ($\gamma = 1$) injection spectrum 
 (empty squares in Fig.~\ref{fig:fluxnu2})
 interactions with IRB photons generate almost as
 many cosmogenic neutrinos as interactions on MBR. The peak of the
 total neutrino energy spectrum from interactions in the MBR and in IRB
 is shifted to lower energy by a small
 amount (from 3$\times$10$^{17}$ eV to about 2$\times$10$^{17}$ eV).
 The distribution extends to lower neutrino energies by more than
 half a decade.
 
 For steeper spectra ($\gamma$ = 1.5) the contribution
 of IRB-generated neutrinos is more significant and the resulting 
 flux is almost an order of magnitude larger.
 The magnitude of the flux at the peak of the spectrum 
 is $\sim 10^{-16}$ cm$^{-3}$s$^{-1}$sr$^{-1}$ and is higher than
 that of the MBR-generated  neutrinos by a factor of $\sim 3$.
 
 In both cases there is no increase of the neutrino flux at energies 
 exceeding 10$^{19}$ eV. The influence of increased cosmological
 evolution of the cosmic ray sources is practically the same 
 as in the case of MBR-generated neutrinos alone. 
 In this context, we note that the cosmological evolution
 of target IR photon density is slower that of the MBR. 

 For comparison, the shaded area in Fig.~\ref{fig:fluxnu2} shows the 
 upper bound
 on the astrophysical neutrino spectra given in Ref.~\cite{WB99}.
 The lower edge is the bound in absence of cosmic ray source
 cosmological evolution, and the upper edge is for $(1 + z)^3$ evolution. 

 The inclusion of the proton-IRB interactions somewhat reverses the
 trend of the injection spectrum dependence of the cosmogenic neutrino
 flux. Without including such interactions, steeper injection spectra
 lead to smaller cosmogenic neutrino fluxes; with the inclusion of the
 contribution to the neutrino flux from interactions of IRB photons with
 relatively lower energy protons, steeper cosmic ray spectra generate
 higher neutrino fluxes. The reason is that we normalize the cosmic ray
 injection spectrum at 10$^{19}$ eV, which is now in the middle of 
 the energy range of the interacting cosmic rays. One can see in
 Fig.~\ref{fig:frac} the dominance of the interactions in the MBR 
 of cosmic rays of energy above 10$^{19}$ at all redshifts higher than 1.
 For steeper cosmic ray injection spectra the number of such 
 particles is decreased while that of cosmic rays below 10$^{19}$ eV,
 that interact in the IRB, is significantly increased. 
 
 Because of the lower average energy of the IRB-generated neutrinos,
 the spectra are shifted and their detectability is lower than that
 of MBR generated photons. This is because the neutrino-nucleon
 cross section rises monotonically with energy. Table~\ref{tab:rates}
 shows the shower rates of $\nu_e$ and $\bar{\nu}_e$  CC (charged current)
 interactions per km$^3$yr of water detector for cosmogenic
 neutrinos generated by interactions with  MBR and IRB photons.
 These rates are the products of the neutrino 
 cross section times the neutrino flux integrated above E$_\tu{sh}$.
 We assume that the total neutrino energy is transfered to the shower
 initiated by its CC interaction. The calculation of event rates
 of CC and NC interactions of muon and tau neutrinos are much more
 difficult and require Monte Carlo models of particular 
 experiments. 

 The second column of Table~\ref{tab:rates} corresponds to the numbers
 given in a
 similar table in Ref.~\cite{ESS01}. The numbers can not be directly
 compared because of the different cosmologies ($\Omega_M$ = 1 in
 Ref.~\cite{ESS01} and $\Omega_M$ = 0.3 here) and cosmological
 evolutions of the cosmic ray sources used. The $\Omega_M$ = 0.3
 cosmology increases the neutrino rates by about 70\%.

\begin{table}
\caption {Rates per km\protect$^3$ water per year of showers
 above different energy  generated
 by different types of neutrino interactions for cosmic ray
 power  density \protect${\cal P}_0$ at $z$=0 of
 1.4\protect$\times {\rm 10}^{31}$ W Mpc$^{-3}$
 and {\em fast} cosmological evolution
 for homogeneously distributed cosmic ray sources (see text).  
\label{tab:rates} }
\medskip
\begin{tabular}{| r | c c | c c |}
\hline
 log\protect$_{10}$ E\protect$_\tu{sh}$
 & \multicolumn{2}{c |}{\protect$\gamma$=1}
 & \multicolumn{2}{c |}{\protect$\gamma$=1.5}\\ \cline{2-5}
 (GeV) \protect$>$ & MBR & IRB & MBR & IRB \\ \hline
 6 & 0.092 & 0.021 & 0.078 & 0.085 \\
 7 & 0.088 & 0.019 & 0.072 & 0.072 \\
 8 & 0.079 & 0.010 & 0.063 & 0.030 \\
 9 & 0.044 & 0.001 & 0.027 & 0.001 \\
10 & 0.008 & 0.000 & 0.003 & 0.000 \\
\hline
\end{tabular}
\end{table}

 In the $\gamma$ = 1.0 injection case the rates from MBR neutrinos
 are higher by factors above 4 at all shower thresholds, while in
 the $\gamma$ = 1.5 case the IRB rates are higher or similar for 
 shower thresholds below 10$^8$ GeV. Above that energy MBR
 neutrinos generate higher rate. Note that IRB neutrinos
 do not contribute at all to the shower rates above 10$^{9}$ GeV.

 The total shower rate for E$_\tu{sh} > $ 10$^6$ GeV is higher than
 the MBR only rate by about 20\% in the $\gamma$=1.0 case, while 
 in the $\gamma$ = 1.5 case it more than doubles the detection rate.
     
 Most of the contemporary fits of the injection spectrum of the highest
 energy cosmic rays confirm the conclusion of Ref.~\cite{BGG05} that 
 it is steeper than $E^{-2}$. The spectrum derived by these authors 
 is E$^{-2.7}$ with a significant flattening at about 10$^{18}$ eV, 
 which could be
 explained with different effects, see e.g. Ref.~\cite{AB05}.
 The shape of the spectrum may be accounted for in this model 
 as a result of the $p \gamma \rightarrow e^+ e^-$ process \cite{BL70}
 as discussed in 
 Ref.~\cite{BG88}. This pair production process creates a dip at
 about 10$^{19}$ eV and a slight excess at the transition from
 pair production to purely adiabatic proton energy loss at
 about 10$^{18}$ eV.
 This fit does not require a strong cosmological evolution
 of the cosmic ray sources, but can accommodate a mild one
 $\propto (1 + z)^m$ with $m \le 3$~\cite{DDMTS05}. In the case
 of flatter injection spectrum the
 pair production dip is well fit also by $m$ = 4.
\begin{figure}
\centerline{\includegraphics[width=83truemm]{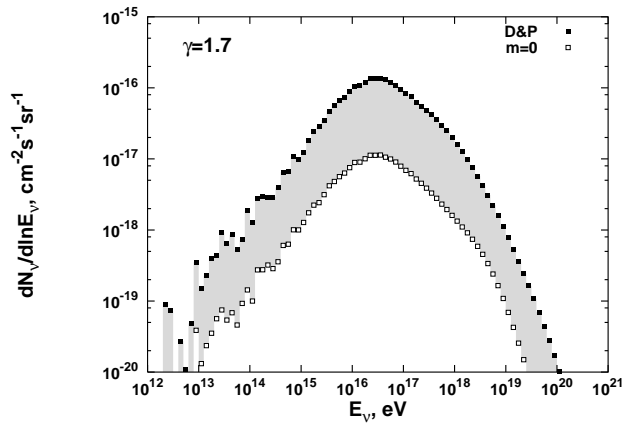}}  
\caption{Total muon $\nu_{\mu} + \bar{\nu}_{\mu}$ fluxes
 for $\gamma$ = 1.7 and evolution from Ref.~\protect~\cite{DP90} 
source evolution (full squares) and no evolution (empty squares). }
\label{fig:BGG}
\end{figure} 
 Figure~\ref{fig:BGG} shows the spectra of the cosmogenic neutrinos 
 from interactions with MBR and IRB photons assuming a steep injection
 spectrum with $\gamma$ = 1.7 and (1) no evolution with ($ m$ = 0)
 and (2) evolution according to Ref.~\cite{DP90}. The difference
 between the two neutrino spectra is
 significant; the peak values are 10$^{-17}$(1.5$\times$10$^{-16}$)
 cm$^{-2}$s$^{-1}$sr$^{-1}$ for without and with evolution.
 The addition of the IRB
 component brings these spectra into the range of detectability,
 especially in the case of mild cosmological evolution.  

\section{Discussion and Conclusions}
 
 The evolution models of SMS05 do not give the highest
 IRB-generated neutrino flux.
 We compared the IRB photon  density in this model with the
 models of Refs.~\cite{F01,KMH02}. Both of these models have
 higher IRB density in the relevant energy range between 3$\times$10$^{-3}$
 and 1 eV. The total IRB densities in the this range are 1.27~\cite{F01}
 and 1.12~\cite{KMH02} compared with the density of 1.03 used here.
 In addition, Ref.~\cite{KMH02} shows somewhat faster cosmological
 evolution. The use of any of these models would have increased somewhat
 the calculated flux of cosmogenic neutrinos. The uncertainty in the
 IRB flux is of the order of 30\%~\cite{SMS05}, while
 the biggest uncertainty in this calculation is in the UHECR flux
 and its cosmological evolution.
 
  The IRB contribution to the total cosmogenic neutrinos flux
 can be slightly increased as protons of energy below
 10$^{18}$ eV can interact with IRB photons and generate lower
 energy neutrinos. If such interactions were included the IRB
 spectrum would be wider than the MBR one, especially at 
 energies below 10$^{16}$ eV.

 It is difficult to compare our results with those of
 Refs.~\cite{BMM04,BK05} because of the different 
 astrophysical input in these calculations. Qualitatively
 the results of these calculations are similar to ours
 and certainly agree within a factor of 2.

 In conclusion, we calculated the flux of cosmogenic neutrinos
 from interactions of UHECR protons with IRB photons using the 
 recent calculations of IR photon spectra densities as a function of
 redshift by SMS05. Our calculations show
 that UHECR interactions with IRB photons produce a significant flux 
 of cosmogenic neutrinos, one which is
 comparable to interactions with MBR photons. This is especially true in
 the case of assumed   
 steep injection spectra of the ultrahigh energy cosmic rays. 
 The total neutrino event rates at energies above 1 PeV increase
 by more than a factor of 2 in the case of injection spectra with
 $\gamma$ = 1.5. Because of the much lower mean free path of
 protons above 10$^{20}$ eV in the MBR interactions
 with IRB photons do not increase the higher energy end of the 
 cosmogenic neutrino spectrum.
 The total cosmogenic fluxes, however, are still not detectable 
 with conventional neutrino telescopes such as 
 IceCube~\cite{IceCube} or the European km$^3$ telescope~\cite{km3}.
 A reliable detection is only expected from radio~\cite{radio}
 and acoustic neutrino detectors.

 {\bf Acknowledgments} We thank Tanja Kneiske for sharing with us
 the tables of cosmological evolution of the IR background from
 Ref. \cite{KMH02}. This work was supported in part by NASA grants 
 ATP03-0000-0057 and  ATP03-0000-0080.

\end{document}